\documentstyle[twocolumn,aps,prb,epsf]{revtex}

\begin{document}

\twocolumn[\hsize\textwidth\columnwidth\hsize\csname
@twocolumnfalse\endcsname

\title{Tight Binding Molecular Dynamics Studies of Boron Assisted
Nanotube Growth}

\author{E.~Hern\'{a}ndez \thanks{To whom correspondence should be addressed.
email: ehe@icmab.es}
and P.~Ordej\'{o}n}
\address{Institut de Ci\`{e}ncia de Materials 
de Barcelona - CSIC,
Campus de la Universitat Aut\`{o}noma de Barcelona, 08193 Bellaterra, Barcelona,
Spain}

\author{I.~Boustani}
\address{Bergische Universit\"at - Gesamthochschule Wuppertal, 
FB 9 - Theoretische Chemie, Gau\ss{}stra\ss{}e 20, D-42097 Wuppertal, Germany}

\author{A.~Rubio and J.A.~Alonso}
\address{Departamento de F\'{\i}sica Te\'{o}rica, Universidad de 
Valladolid, 47011 Valladolid, Spain}

\date{\today}

\maketitle

\begin{abstract}
In this paper we report a theoretical study of the effects of the presence
of boron in growing carbon nanotubes. We employ a well established Tight
Binding model to describe the interactions responsible for the energetics 
of these systems, combined with the Molecular Dynamics simulation
technique and Structural Relaxation calculations. We find, in agreement with 
the previous theoretical/experimental work of 
Blase {\em et al.\/} 
\mbox{[}{\em Phys. Rev. Lett.\/} {\bf 83}, 5078 (1999)\mbox{]}, 
that boron favors (n,0) (zig-zag) tubular structures
over (n,n) (arm-chair) ones by stabilizing the zig-zag edge. Furthermore, 
it is shown that boron has the effect of delaying the tube closure 
process, a fact which could explain the improved aspect ratio experimentally
observed in nanotubes synthesized in the presence of boron. Our 
dynamical simulations lead us to propose a mechanism through which this 
extension of the closure time can be explained.
\end{abstract}


]

\narrowtext

\section{Introduction}
\label{sec:introduction}

The discovery of carbon nanotubes by Iijima in 1991~\cite{iijima} has marked
the starting point of a scientific
revolution~\cite{dresselhauss_book,ebbesen_book,ajayan:ebbesen_rpp,ajayan_cr,mauro_rev}.
This discovery has opened a whole new perspective in the nanoscopic
regime of Materials Science and Engineering. Their
mechanical~\cite{treacy,zettl,krishnan,lieber,hernandez,lum,yakobson,buongiorno}, 
electrical~\cite{hamada,wildoer} and 
magnetic~\cite{lu,knechtel,deHeer} properties provide ample opportunity
for the fabrication of nano-scale devices. Indeed some such devices have
already been reported in the literature~\cite{stm,field,devices1,devices2}.
Since Iijima's, discovery nanotubes of other chemical composition have also 
been synthesized, such as $\mbox{B}_x\mbox{C}_y\mbox{N}_z$ composite
nanotubes~\cite{bcn1,bcn2,bcn3,bcn4,bcn5}, the so-called inorganic
nanotubes consisting of layers of $\mbox{MoS}_2$ or 
$\mbox{WS}_2$~\cite{tenne1,tenne2,tenne3}, or $\mbox{NiCl}_2$ 
nanotubes\cite{sloan}. All these compounds have phases which consist of
layered structures, and this has lead to the prediction that other 
materials also capable of crystallizing in layered structures can in principle
produce nanotubes. Indeed, theoretical arguments have been 
presented in the literature for the viability of 
BN~\cite{corkill_bn,blase_bn}, $\mbox{BC}_3$~\cite{miyamoto_bc3},
$\mbox{BC}_2\mbox{N}$~\cite{miyamoto_bc2n}, GaN~\cite{lee:lee}, 
B~\cite{boustani}, 
GaSe~\cite{cote} and P~\cite{seifert:hernandez} nanotubes. Interestingly,
nanotubular structures can also be constructed from biochemical compounts,
such as peptides, as demonstrated experimentally by 
Ghadiri {\em et al.\/}~\cite{ghadiri} and theoretically by
Carloni {\em et al.\/}~\cite{carloni}.

The tubes first detected by Iijima~\cite{iijima} were multi-wall 
nanotubes (MWCNT's), {\em i.e.\/} concentric shells of cylindrical shape,
in which each shell is separated from the next by approximately the
same distance as the interlayer spacing in graphite. Each shell can be 
characterized by a pair of indices, (n,m), which determine how the 
folding of the graphene sheet must be carried out in order to obtain 
the shell. The shells are usually classified into three different
types: (n,0) or {\em zig-zag\/} shells, (n,n) or {\em arm-chair\/}
shells, and {\em chiral\/} shells, of indices (n,m) where
$\mbox{n} > \mbox{m} > \mbox{0}$. Zig-zag and arm-chair shells are
said to be achiral (they can be superimposed onto their mirror images).
The ordering of the shells in a MWCNT is usually
{\em turbostratic\/}, {\em i.e.\/} the pattern of atomic arrangement may
vary (and in general does vary) from one shell to the next, or in 
other words, different shells usually have different 
chiralities~\cite{ajayan:ebbesen_rpp}.

After the discovery of MWCNT's, a procedure for synthesizing 
single-wall nanotubes (SWCNT's) was found~\cite{thess}. These tubes are
found to aggregate into bundles or ropes, and their diameter distribution
peaks at around 1.4~nm, although more recently this distribution has been
found to vary according to the synthesis conditions~\cite{bandow}. 
The production of
SWNT's has allowed the experimental corroboration~\cite{wildoer} of a 
theoretical prediction made by Hamada and coworkers~\cite{hamada} soon after 
the discovery of MWCNT's, that the electrical conductivity properties 
of SWCNT's are dependent on the (n,m) indices. This prediction stated
that SWCNT's can be either metallic or semi-conducting; if the indices 
(n,m) of a nanotube obey the relation $\mbox{n}-\mbox{m} = 3 \mbox{q}$
(q = 0, 1, 2, 3, \ldots), then the tube is metallic, otherwise the tube
is semi-conducting.
This dependence of the electric characteristics of SWCNT's upon their
structure has raised an interest in the possibility of devising new
synthetic methods that allowed a structural selection of nanotubes,
not only according to their diameter, but also to their chirality. A first
step in this direction was achieved by Redlich {\em et al.\/}~\cite{redlich},
Carroll {\em et al.\/}~\cite{carroll}, 
and by Terrones {\em et al.\/}~\cite{mauro_apl}, who, by adding a certain
amount of boron during the synthesis, obtained boron doped MWCNT's which,
interestingly, have an improved crystallinity and a larger
aspect ratio (quotient of length to diameter) with respect to tubes 
obtained in the absence of boron. It was shown that boron appears mostly
in the form of clusters associated with the tips of the nanotubes.
But most importantly, a recent combined experimental and theoretical
study by Blase and coworkers~\cite{blase:terrones} has demonstrated that
the MWCNT's thus obtained consist mostly of zig-zag shells. This study also
sheds some light on the role played by boron in favoring the zig-zag
structure over others, and tries to explain the larger aspect ratio
observed in boron assisted synthesis. This latter issue, though, was 
addressed by First-Principles~(FP) Density Functional Theory (DFT)
Molecular Dynamics simulations, and the
large computational costs of this technique prevented Blase and 
coworkers from pursuing a detailed enough study which clarified completely
this question.

In this paper we address the problem of boron assisted nanotube growth
using a Tight Binding model~\cite{tb_review}. Tight Binding~(TB) is
an approximate method which nevertheless is capable of providing extremely
accurate results in favorable systems. Its main advantage with respect
to FP methodologies is its comparatively low computational cost, which 
often allows a more extensive study than is practical or even possible
with higher levels of theory. In this work we have used the 
Density-Functional Tight Binding (DFTB) model due to Porezag 
{\em et al\/}~\cite{porezag}, about which we give more details in 
Section~\ref{sec:computational}. This model has proved to be very
accurate for carbon based systems. We have used DFTB to perform a 
series of static and dynamical simulations of SWCNT's, with and without
boron present, with the aim of understanding the effects of the 
presence of boron on the structural properties of the resulting NT's.
The structure of this paper is as follows: in Section~\ref{sec:computational}
we describe briefly the DFTB model, and provide previous examples of its
successes in order to justify its use here. We also describe the calculations
which are reported in the remaining of the paper. Section~\ref{sec:results}
is devoted to a discussion of our simulation results, and we summarize 
our conclusions in Section~\ref{sec:concs}.

\section{Computational Details}
\label{sec:computational}
\subsection{Model}
\label{sub:model}

DFTB is is a 
non-orthogonal Tight Binding scheme in which a parametrisation is 
constructed directly from DFT calculations
using atomic-like orbitals in the basis set, and adopting a two-center
approximation for the Hamiltonian matrix elements. For more details on the
parametrisation used here the reader should consult 
references \cite{porezag,widany}. The DFTB scheme has proved to be 
extremely successful in the modeling of carbon-based systems, in 
particular carbon clusters and nanostructures. Fowler {\em et al.\/}
\cite{fowler} have used it 
to analyze the energetic ordering of all 426 cage structures 
containing 5, 6 and 7-membered rings in $\mbox{C}_{40}$.
Ayuela {\em et al\/.}~\cite{ayuela}
 have found, using DFTB and other five semi-empirical
methods, a heptagon-containing isomer of $\mbox{C}_{62}$ which
was predicted to be more stable than any of the other 2385
classical fullerene isomers ({\em i.e.} isomers containing only pentagons
and hexagons). 
DFTB has also been used to determine the mechanical properties of 
single-wall C, BN and some $\mbox{B}_x\mbox{C}_y\mbox{N}_z$ 
nanotubes \cite{hernandez}, providing results which are in excellent 
agreement with the available experimental data. More recently DFTB has
also been used to study the structural, mechanical and electronic
properties of a novel family of laminar carbon structures known
as {\em Haeckelites\/}, as well as those of their tubular
counterparts \cite{haeckelites}. Haeckelites consist of pentagons
and heptagons in equal number, with an arbitrary number of hexagons.
The suitability of DFTB for performing
Molecular Dynamics simulations in carbon-based systems has been most
recently demonstrated by Fugaciu {\em et al.\/}~\cite{fugaciu}, who have
used it to study the conversion of diamond nanoparticles to concentric
shell fullerenes.
The many examples of the use of DFTB in the context of carbon based systems
and the accuracy of the results reported give us confidence in the 
reliability of this theoretical model. 

Like in other Tight Binding models \cite{tb_review}, in DFTB the total 
energy is calculated
as the sum of two contributions: the {\em band structure\/} contribution,
which is mostly attractive, and the repulsive {\em pair-potential\/} 
contribution, which accounts for the core-core repulsion and 
the double-counting of the electron-electron interaction which is implicit in
the band structure term. The band structure
energy is calculated by straight forward diagonalisation of the Hamiltonian,
summing the eigen-values of the occupied states weighted according to their
occupation numbers, {\em i.e\/}
\begin{eqnarray}
E_{bs} = 2 \sum_n f_n \epsilon_n,
\label{eq:energy}
\end{eqnarray}
where $f_n$ is the population of state {\em n\/} (which in our case is equal
to 1 for occupied states, and to zero for unoccupied, {\em i.e.\/} we have
assumed 0~K electronic temperature), and $\epsilon_n$ is the
eigen-value of state {\em n\/}. The factor of two accounts for the degeneracy
of spin. The band structure contribution to the atomic force 
$\mbox{\bf F}_i$ is then calculated from the Hellmann-Feynman theorem:
\begin{eqnarray}
\nabla_i\, \epsilon_n = 2 f_n \mbox{\bf C}_n^\dag\left( \nabla_i\, \mbox{\bf H}
               - \epsilon_n \nabla_i\, \mbox{\bf S} \right) \mbox{\bf C}_n,
\label{eq:force}
\end{eqnarray}
where $\mbox{C}_n$ is the vector representation of eigen-state {\em n\/},
{\bf H} is the Hamiltonian matrix, and {\bf S} is the overlap.
The repulsive 
pair potential energy and force contributions are trivially added to 
equations \ref{eq:energy} and \ref{eq:force}, respectively.

\subsection{Calculations}
\label{sub:calculations}

We have performed two types of calculations: a) structural
relaxation calculations, in which, using the Conjugate Gradients (CG)
technique \cite{numericalrecipes}, the positions of the atoms in a 
system are displaced until a minimum in the potential energy hyper-surface
is found; and b) molecular dynamics (MD) simulations, which we have employed
to study the time evolution of the systems considered here at a range
of temperatures. More specifically we have performed 
canonical ensemble molecular dynamics calculations, in which the conserved
quantities are the number of atoms, the volume and the average
temperature (NVT-MD). We have implemented the NVT-MD algorithm of 
Nos\'{e} as modified by Hoover \cite{nose,hoover}. In this algorithm
the physical system of interest is extended by the addition of an
extra degree of freedom, playing the role of a thermostat, which interacts
with the physical system in such a way as to fix the average temperature.
In this algorithm the mass associated with the thermostat 
is somewhat arbitrary, as it does not affect the value of 
the average temperature, only the size of the instantaneous temperature
fluctuations. We have chosen this mass to be equal to the total mass of
the physical system.
We have used the implementation of the Nos\'{e}-Hoover algorithm 
described by Frenkel and Smit \cite{frenkel:smit}. 
Although the total energy is not conserved in NVT-MD, there is a 
conserved quantity which can be monitored to ensure the correctness of
the implementation. This magnitude plays the role of total energy of the
extended system. In our simulations we have used a time-step of 1 fs,
which has proved to be sufficiently small to maintain the conserved
quantity oscillations smaller than 1 part in 10000 during the length of
the simulations, with no appreciable drift.

Our studies have covered a range of temperatures, namely 1000, 2000, 2500
and 3000 K. At each temperature the dynamics of the system were monitored
during at least 10~ps. The simulations were carried
out sequentially, using the coordinates and velocities at the end of a run
at a given temperature as a seed for the simulation at the next higher
temperature. The transition from one temperature to another was made by
gradually increasing the temperature of the thermostat at a rate 
of 0.5 K/fs. To use a lower heating rate would be prohibitive, in view of the
computational costs involved. Nevertheless, once the thermostat reached the 
desired temperature, the system was allowed to reach equilibrium at the new 
temperature during 1 ps, in order to minimize as much as possible the
effects of such a high heating rate, before performing the simulation at 
the new temperature.

\section{Simulation Results}
\label{sec:results}

\subsection{Carbon nanotubes in the absence of boron}
\label{sub:undoped}

\begin{figure}
\begin{minipage}[t]{3cm}
  \begin{center}
       {\bf \large (a)}
    \leavevmode
    \epsfxsize=3cm
    \epsffile{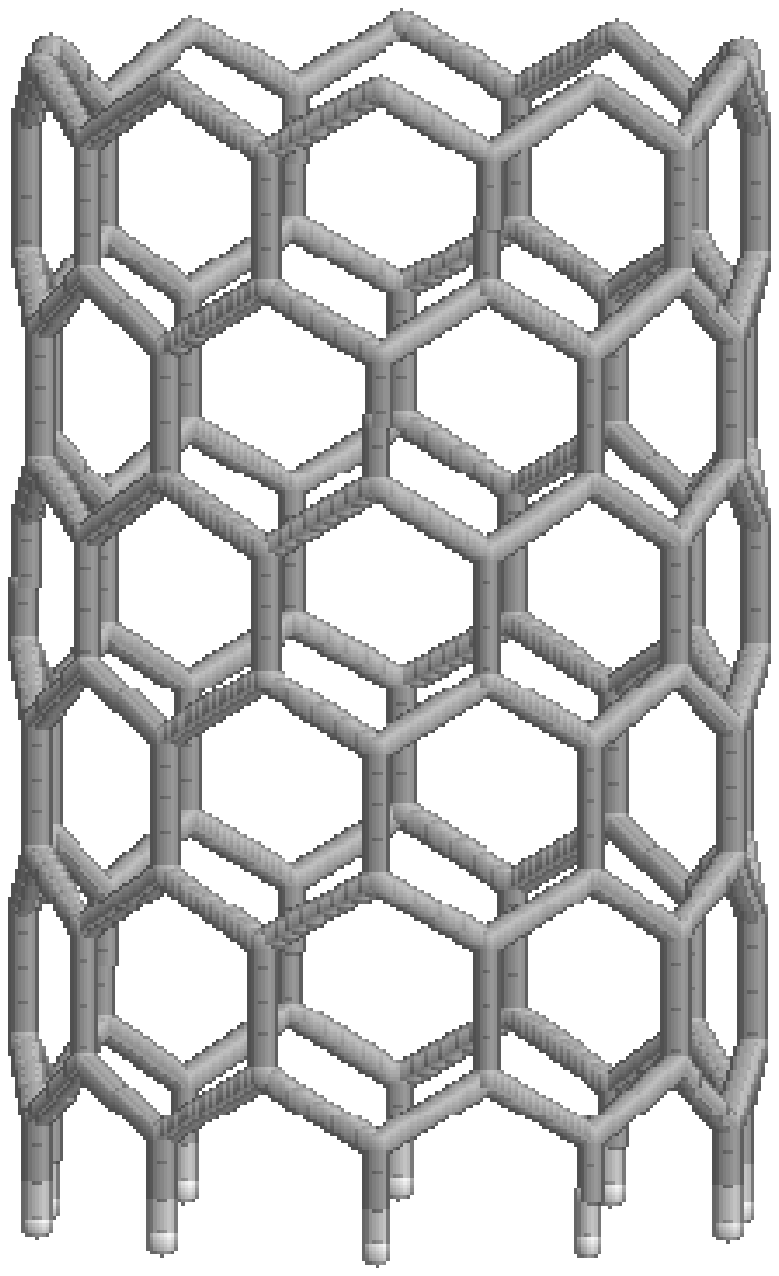}
  \end{center}
\end{minipage}
\begin{minipage}[t]{3cm}
  \begin{center}
       {\bf \large (b)}
    \leavevmode
    \epsfxsize=3cm
    \epsffile{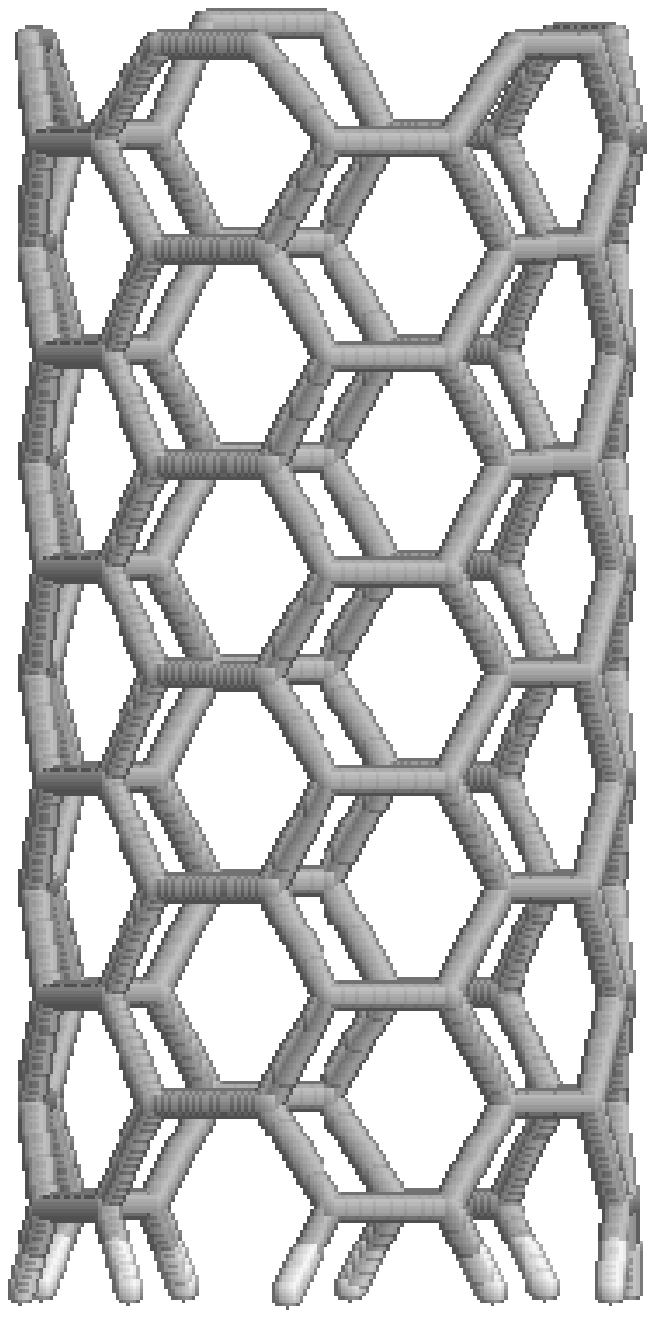}
  \end{center}
\end{minipage}
\caption{The relaxed structures of a (10,0) nanotube (a) and a (5,5)
nanotube (b), used as starting configurations for the Molecular Dynamics
simulations described in the text.}
\label{fig:starting}
\end{figure}

Before addressing the effect of boron on the structure and growth 
mechanism of nanotubes, it is necessary to study the dynamics of 
carbon nanotubes in the absence of boron. In so doing we can also assess
the quality and appropriateness of the model.
In order to study the dynamics of the open ended tubes, we first considered
the systems illustrated in Figure~\ref{fig:starting}. These systems are a 
(10,0) nanotube, with one end saturated by ten H atoms, and a (5,5) nanotube,
also with one end saturated by ten H atoms. Both tubes
consist of 120 C atoms, plus the 10 H atoms. Figure~\ref{fig:starting}
shows the structures obtained after a CG relaxation, which were then used
as starting configurations for the NVT-MD simulations.
To facilitate the analysis of the MD simulations, the 10 H atoms were 
kept fixed throughout the dynamical simulations. While this is an 
approximation, we have chosen a tube section as large as was practical 
in order to minimise the possible effects of having the H atoms frozen 
on the dynamics at the other end of the tube.

\begin{figure}
\begin{minipage}[t]{3cm}
\begin{center}
{\Large (a)}
\leavevmode
\epsfxsize=3cm
\epsffile{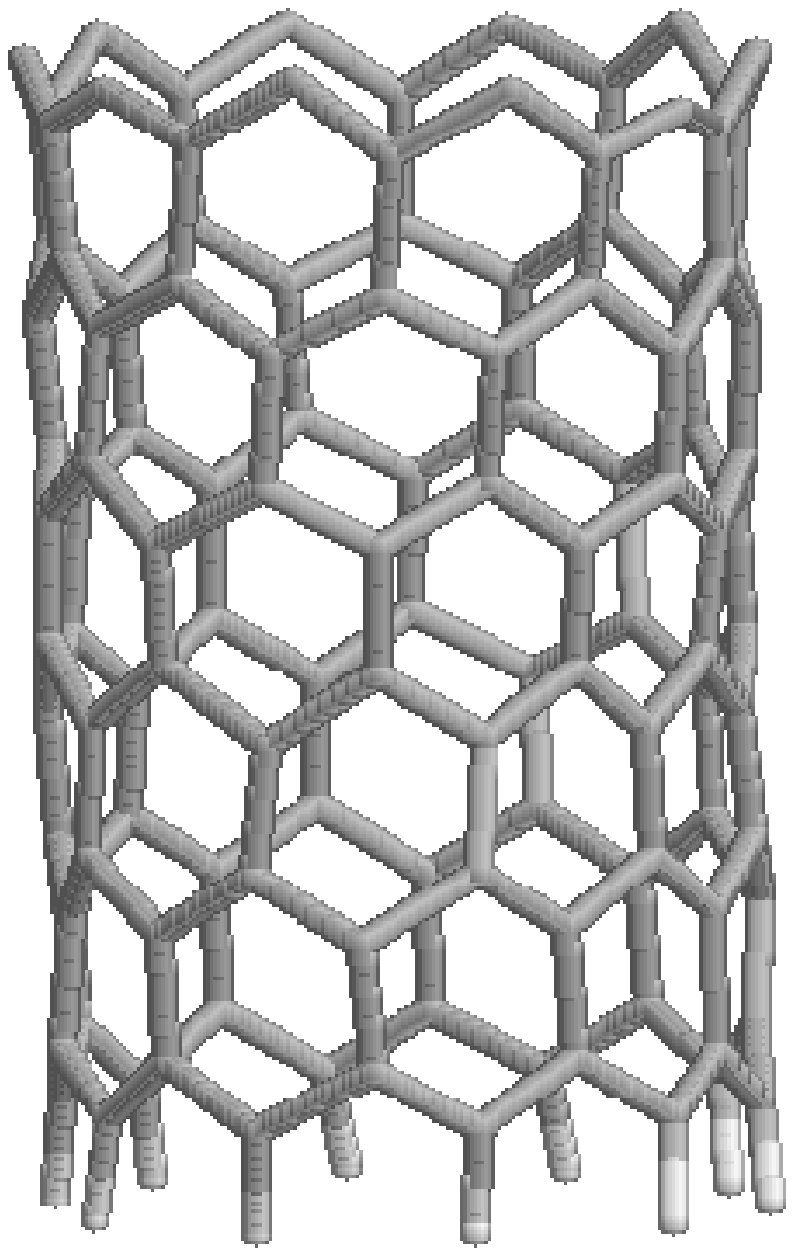}
\end{center}
\end{minipage}
\begin{minipage}[t]{3cm}
\begin{center}
{\Large (b)}
\leavevmode
\epsfxsize=3cm
\epsffile{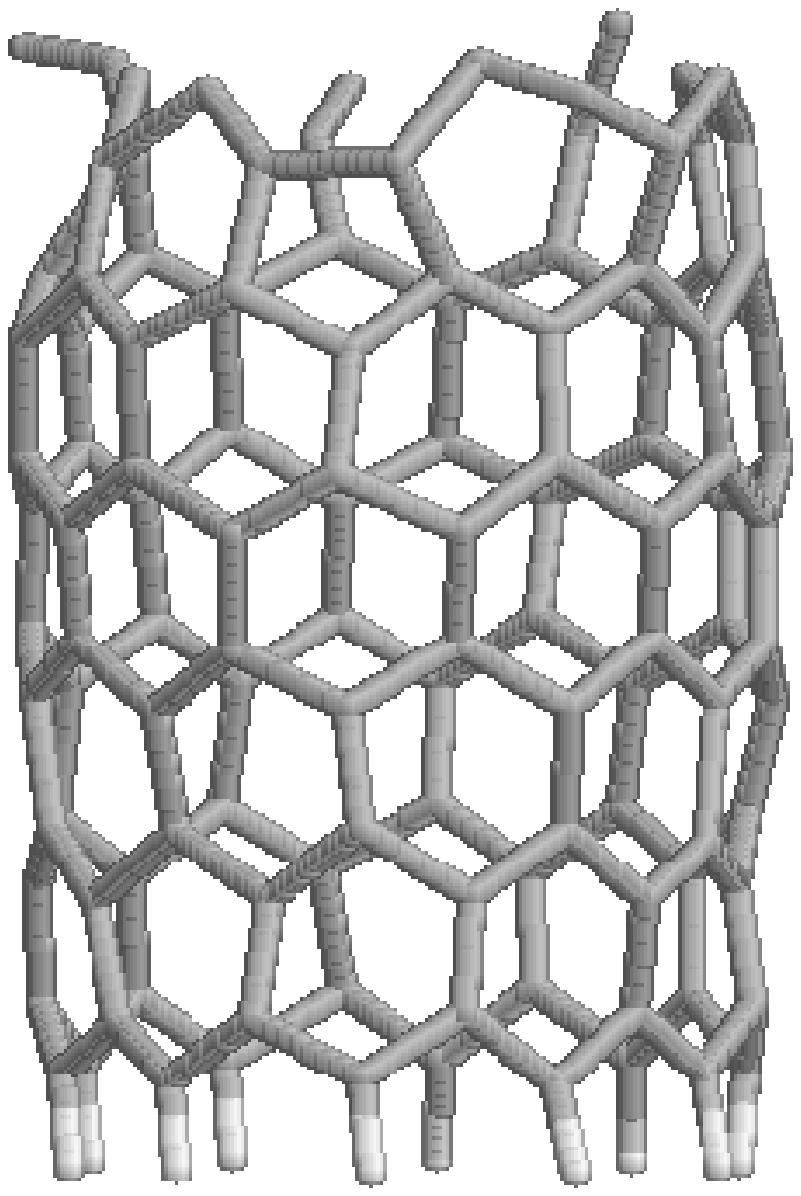}
\end{center}
\end{minipage}
\caption{(a) Final structure of the (10,0) nanotube obtained after 10~ps of
dynamics at 1000~K. (b) Final structure of the (10,0) nanotube obtained
after 10~ps of dynamics at 2000~K.}
\label{fig:1000-2000K}
\end{figure}

We first discuss the case of the (10,0) nanotube.
At the lowest temperature considered (1000 K) the structure of the nanotube
remains unaltered except for the normal thermal oscillations around the
equilibrium atomic positions. All the hexagonal rings present in the initial
structure (Figure~\ref{fig:starting}) retain their identity, as can be
seen in Figure~\ref{fig:1000-2000K}(a), which illustrates the configuration 
attained after 10~ps of MD run at this temperature.
In view of the fact that the simulation of the open nanotube end at 1000 K
failed to produce any structural reordering within this time scale, we 
proceeded to heat the system up to 2000 K. At this temperature, structural
changes begin to appear in the open edges of the nanotube, as can be 
seen in Figure~\ref{fig:1000-2000K}(b). Indeed, some hexagonal 
rings at the edge
have been broken, resulting in chains of carbon atoms, while others have
fused into pentagon-heptagon (5/7) pairs. After 10~ps at 2000 K, two pentagonal
rings can be seen, as well as one heptagon. These non-hexagonal rings are
produced by the fusing of two hexagons, to give a pentagon and a heptagon.
However, we find that heptagons appear to be less stable, and break more
easily into carbon chains that remain attached to the nanotube edge, while
the pentagons remain stable. The presence of these pentagons induces 
curvature in the structure, so the edge of the nanotube bends inwards.

\begin{figure}
\begin{minipage}[t]{3cm}
\begin{center}
{\Large (e)}
\end{center}
\epsfxsize=3cm
\epsffile{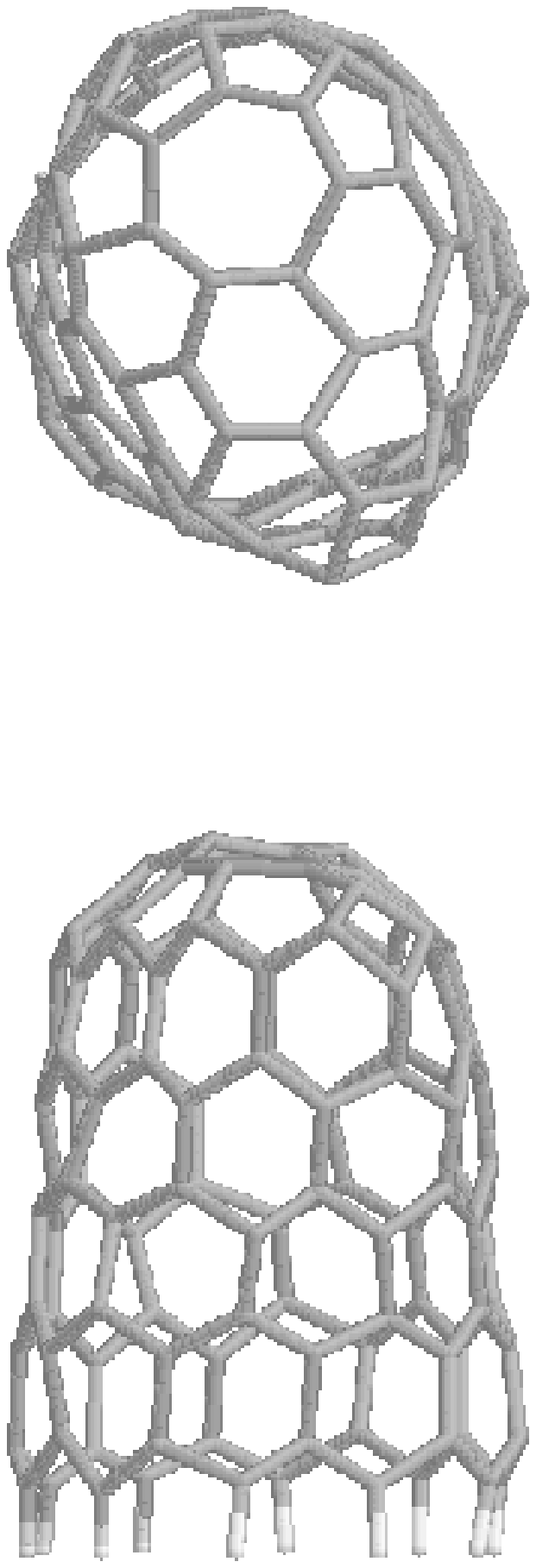}
\end{minipage}
\begin{minipage}[t]{3cm}
\begin{center}
{\Large (f)}
\end{center}
\epsfxsize=3cm
\epsffile{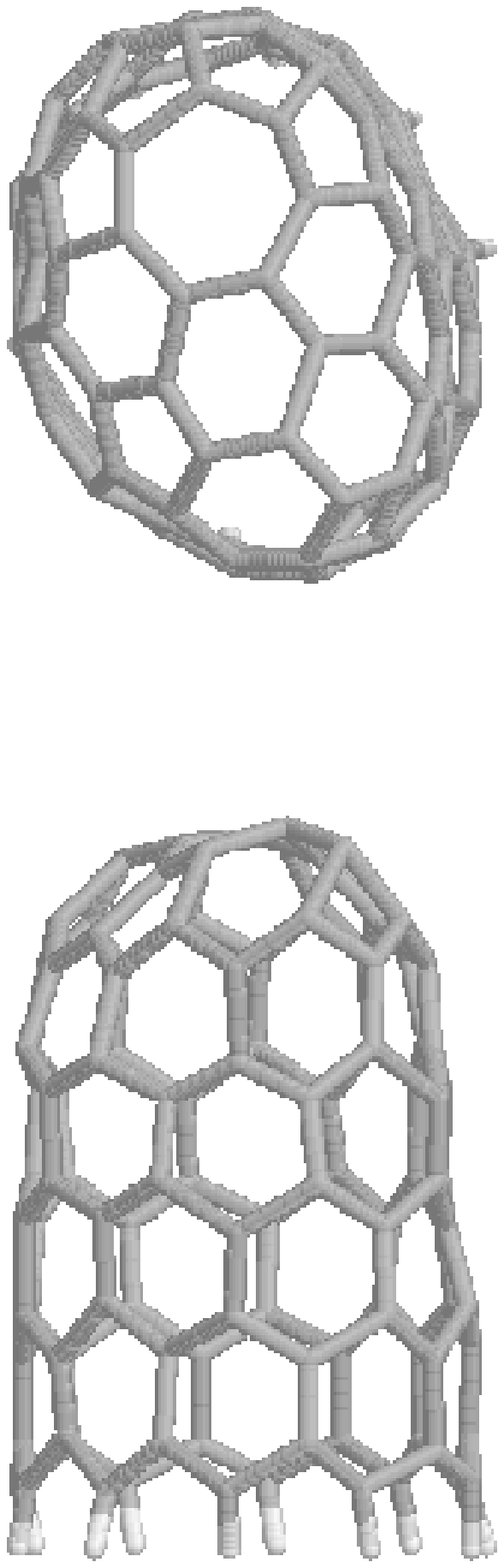}
\end{minipage}
\caption{Different stages of the closure of the (10,0) nanotube at 2500 K.
These structures were observed at approximately (a) 2~ps, (b) 5~ps, (c) 8~ps,
(d) 11~ps, (e) 15~ps and (f) 18~ps.}
\label{fig:dynamics}
\end{figure}

Clearly, a tendency to form a fullerene-like cap is observed, a situation
that is after all energetically more stable than the open edge. Therefore,
we continued the simulation at a temperature of 2500 K. Like before,
the temperature of the thermostat was increased at a rate of 0.5 K/fs,
and once the desired thermostat temperature was achieved the combined
system of nanotube and thermostat was allowed to equilibrate during
1 ps. The dynamics of the system were then monitored during a further 20 ps.
Figure~\ref{fig:dynamics} illustrates six representative configurations
of the system at this temperature. As can be seen, the process of tube
closure initiated at 2000~K is further facilitated at 2500 K. The chains
of carbon atoms which were seen already at 2000~K can now form links
bridging across the open edge. This, combined with the presence of the
pentagonal rings at or close to the edge, has the overall effect of 
bending inwards the nanotube walls. As the simulation proceeds, the 
number of freely oscillating chains is reduced, due to their tendency to 
bond into the dome-like structure being formed. Eventually the tube
closure is completed, and all carbon atoms are three-coordinated in 
an $\mbox{\em sp}^2$ hybridisation fashion. Although the resulting 
closed structure is highly strained, due to the presence of some
small rings (a tetragon can be seen) and to the adjacency of pentagons,
this is certainly a deep local minimum of the potential energy hyper-surface,
as the structure, once closed, remains unaltered (except for thermal
oscillations), even after heating up the system to 3000~K and following 
the dynamics at this temperature for a further 10 ps. Obviously, the 
time-scale for structural re-orderings once the nanotube has closed is 
much larger than the time-scale for the closure itself. This is because
structural changes now must happen via the Stone-Wales \cite{stone:wales}
mechanism, which has a high activation barrier, and therefore occurs
very infrequently \cite{tiffany}.


\begin{figure}
\begin{minipage}[t]{3cm}
\begin{center}
{\Large (a)}
\end{center}
\epsfxsize=3cm
\epsffile{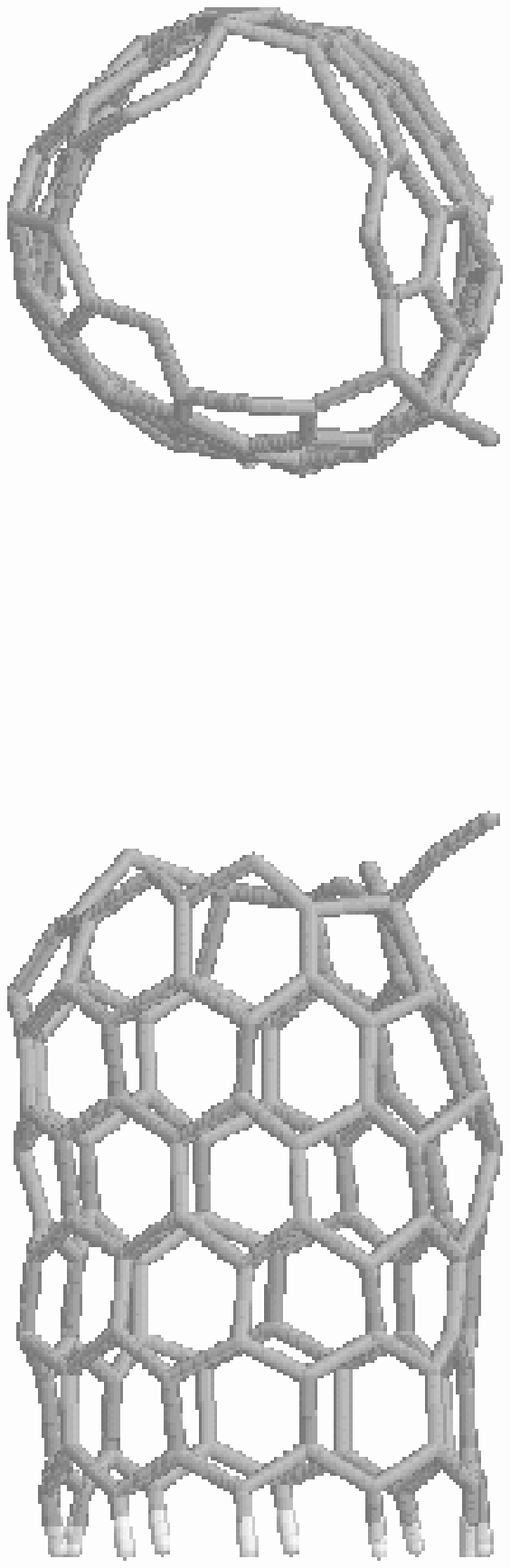}
\end{minipage}
\begin{minipage}[t]{3cm}
\begin{center}
{\Large (b)}
\end{center}
\epsfxsize=3cm
\epsffile{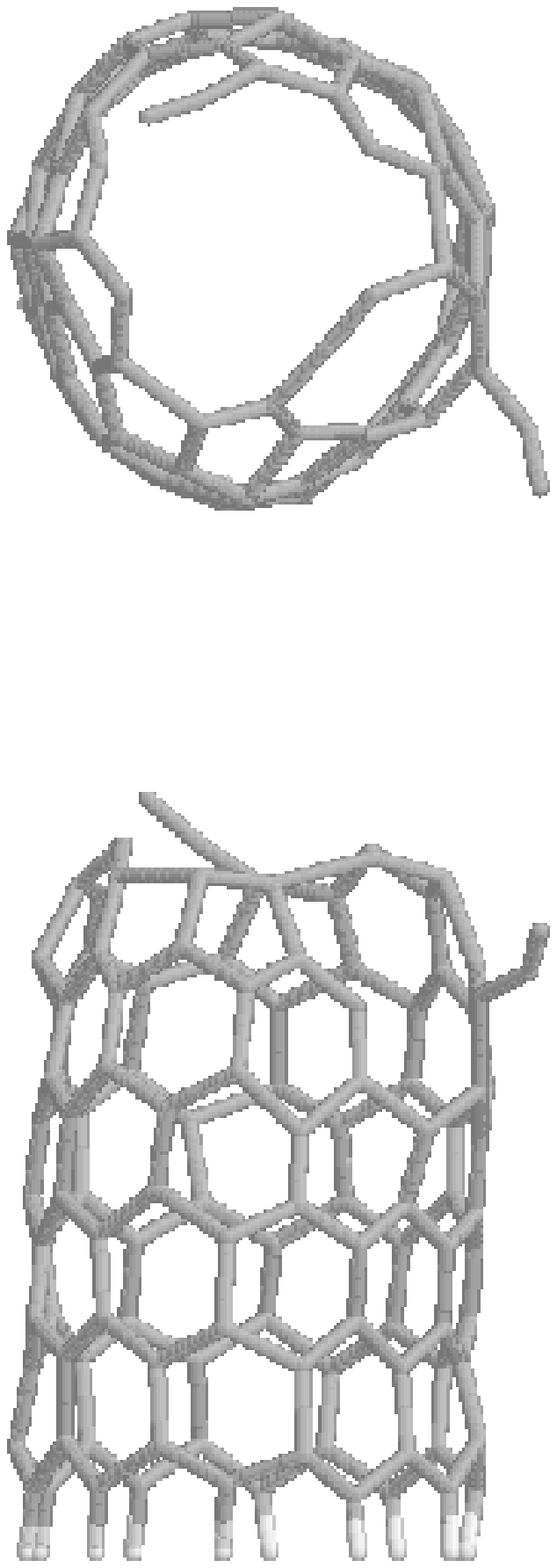}
\end{minipage}
\end{figure}
\begin{figure}
\begin{minipage}[t]{3cm}
\begin{center}
{\Large (c)}
\end{center}
\epsfxsize=3cm
\epsffile{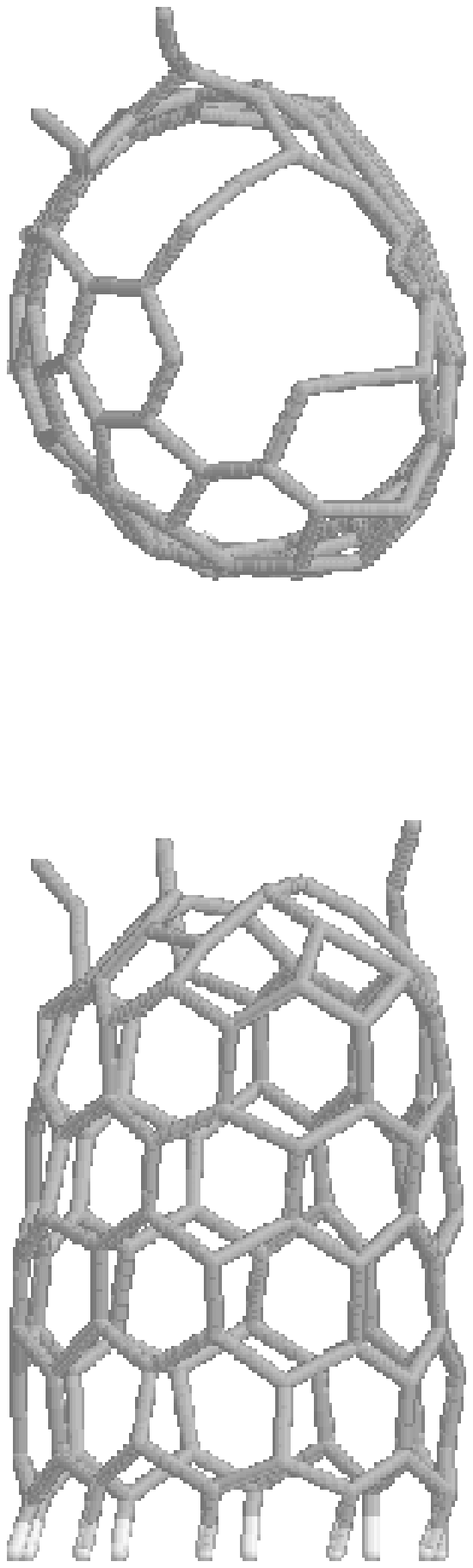}
\end{minipage}
\begin{minipage}[t]{3cm}
\begin{center}
{\Large (d)}
\end{center}
\epsfxsize=3cm
\epsffile{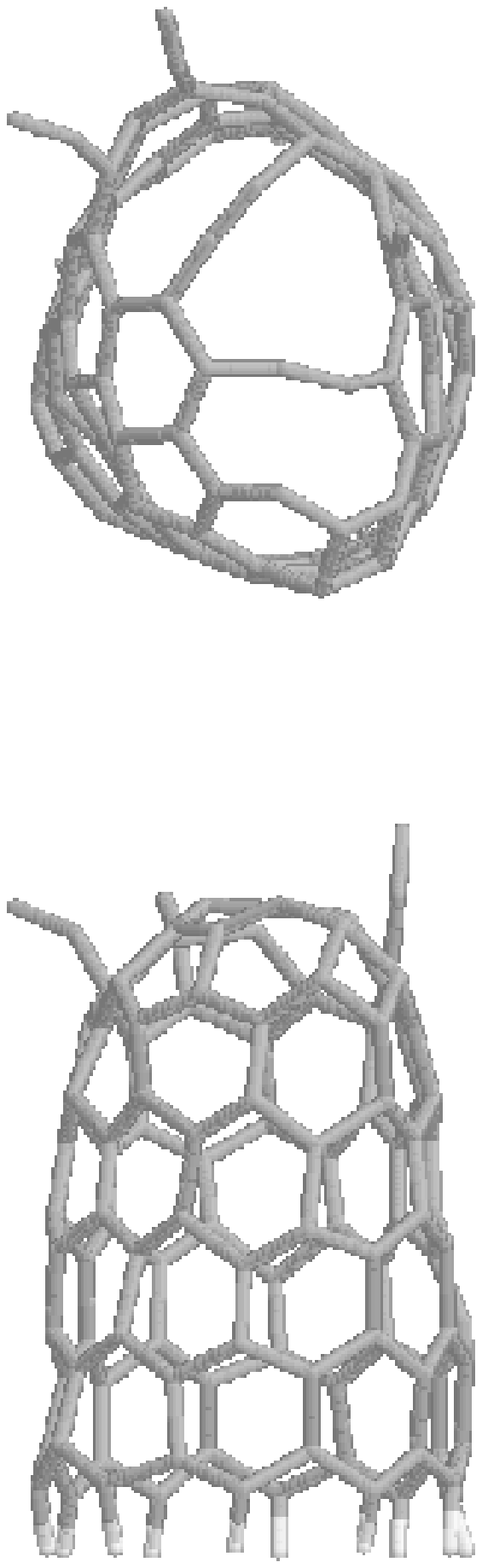}
\end{minipage}
\end{figure}



We now discuss the results obtained from the simulation of the (5,5)
nanotube. During the first 10~ps of dynamics, in which the average temperature
was fixed at 1000~K, one of the edges of a hexagon at the end of the nanotube
is broken, and as a result a carbon chain is formed which oscillates under 
the influence of the thermal motion, but no (5/7)~complexes are 
observed at this temperature. When the temperature is increased to 2000~K
two more hexagons break, but still no (5/7)~pairs are formed.
At 2500~K nearly all edge hexagons have broken, and consequently there
is an abundance of carbon chains. Among these the first two pentagons
can be observed, but there are still no signs of heptagons. This is probably
because the mechanism of pentagon formation in this case is somewhat
different than in the (10,0) nanotube. In the latter, pentagons and
heptagons occur in pairs, resulting from the fusion of two edge hexagons.
Here the pentagons appear from the re-bonding of chains resulting from
the breaking of edge hexagons. The presence of these pentagons induces
curvature in the structure, which will eventually cause the tube closure.
At 3000~K a total of three pentagons are seen, and the structure is 
rapidly evolving towards complete closure, but this only happens beyond
10~ps of dynamics at this temperature.

Our results show unambiguously that the open
edges of nanotubes are unstable, and at sufficiently high temperatures
close spontaneously by forming
half-fullerene caps. The time scales we find for the nanotube closure
are only orientative, as they may vary slightly for different initial
conditions, and certainly they are sensitive to the temperature, but we
can say that above 2000 K, closure occurs within a few tens of ps. This
finding is in agreement with previous first-principles results \cite{charlier}.
At temperatures between 1000 and 2000~K the time scale for closure
(which remains beyond the scope of the simulation times covered here)
could be in the order of hundreds of~ps or even in the ns range, but even
so this seems to us to be too short a time scale for permitting nanotube
growth, and this is presumably why single-wall nanotubes cannot be obtained
without using transition metal nano-particles which act as catalysts.
However, the microscopic details of this catalytic growth process are
as yet far from being understood~\cite{kanzow}.

In order to gain some insight into the experimental observation
that boron doping can result in the production of longer and more
crystalline multi-wall nanotubes, and in particular how it favors the 
zig-zag type tubes over the arm-chair or chiral ones, we have performed
a number of static and dynamical simulations in the (10,0) nanotube system
with different degrees of boron substitution. 

\subsection{Boron doped nanotubes}
\label{sub:doped}

\begin{figure}
\begin{minipage}[t]{6.5cm}
\begin{center}
\leavevmode
\epsfxsize=6.5cm
\epsffile{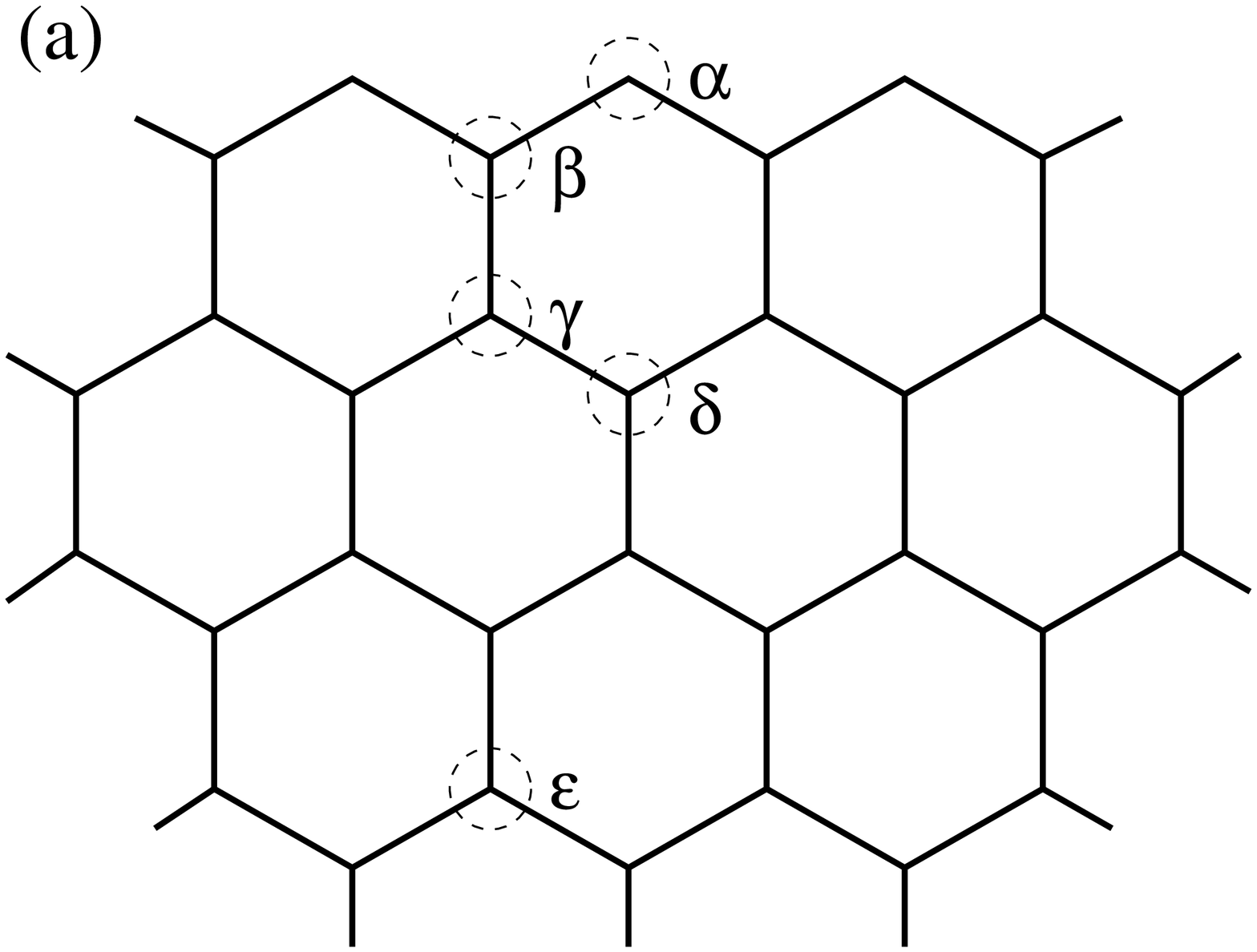}
\end{center}
\end{minipage}
\begin{minipage}[b]{6.5cm}
\begin{center}
\leavevmode
\epsfxsize=6.5cm
\epsffile{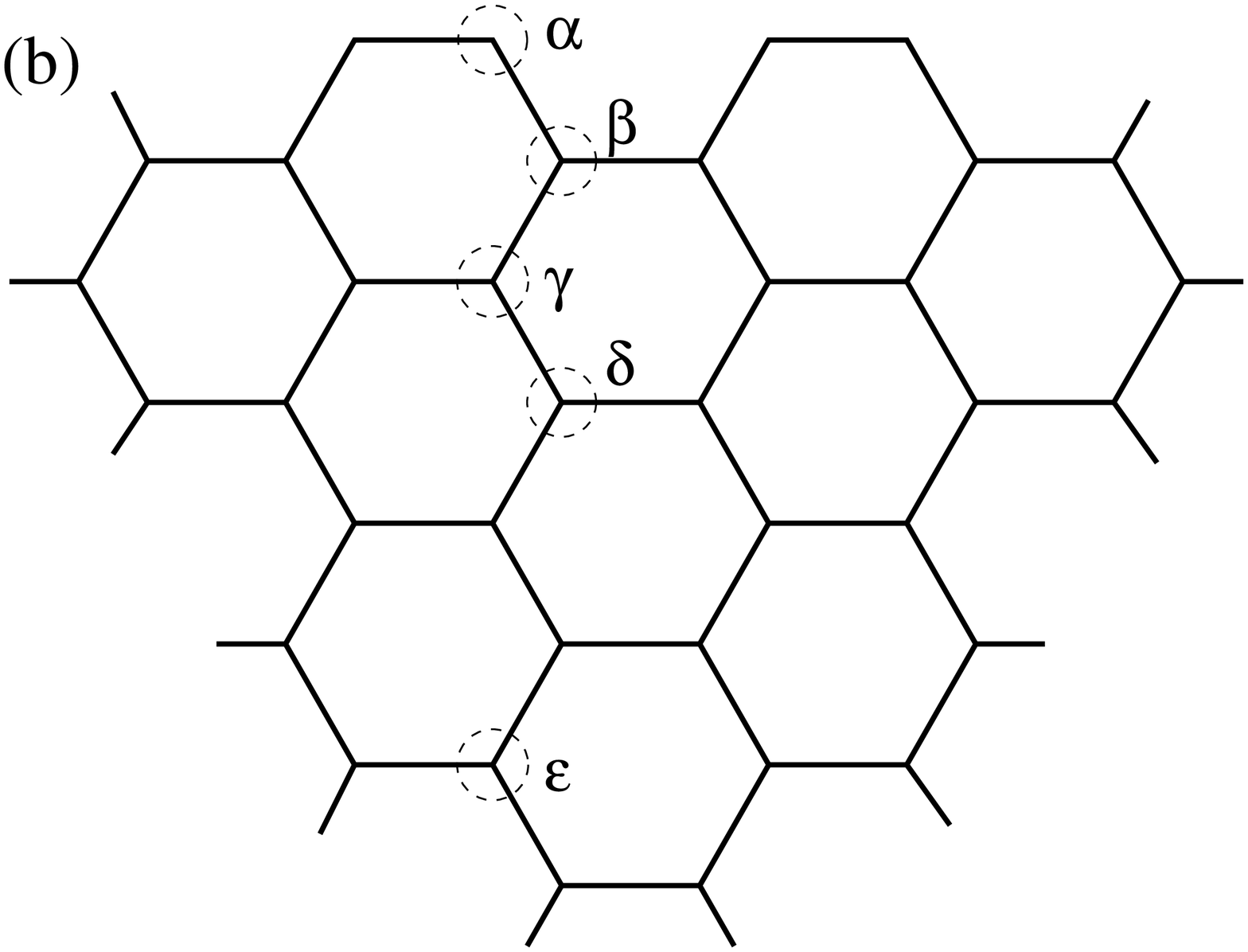}
\end{center}
\end{minipage}
\caption{(a) Schematic view of the boron substitutional sites considered in the
(10,0) nanotube and the zig-zag graphene edge; (b) the substitutional sites
in the (6,6) nanotube and arm-chair graphene edge.}
\label{fig:sites}
\end{figure}

We first address the question as to why boron tends to accumulate at the 
ends of the nanotubes, as has been experimentally 
observed \cite{blase:terrones}. To investigate this site preference, 
we have performed a series of relaxation calculations in which we 
compare the energy of nanotube and flat edge structures with a boron atom 
substituting a carbon atom at different sites. We have considered two
different nanotube structures, one a zig-zag nanotube with indices (10,0)
(the same as illustrated in Figure~\ref{fig:starting}), and a (6,6)
arm-chair nanotube. For each of the two structures we have taken into 
account five possible boron substitutional sites, namely: the
boron atom substitutes at an edge carbon site (site $\alpha$),  
the boron is located at a site one bond away from the edge site (site $\beta$),
two bonds away (site $\gamma$), three (site $\delta$)
and six bonds away from the edge (site $\epsilon$).
These positions are illustrated schematically in Figure~\ref{fig:sites}.

Our results, given in Table~\ref{table:energies}, show that in 
both types of nanotubes the most favorable position
for boron to substitute at is in the vicinity of the edge . However, 
there are marked 
differences between both types of edges. In the zig-zag case, the energy
of the structure with boron substituted in site $\beta$ is 1.4 eV higher
than the structure with boron substituted at the $\alpha$ site, almost the
same as for substitution of boron far (site $\epsilon$)
from the edge (1.41 eV). So, in the
case of the zig-zag edge it is much more favorable for boron to substitute
at the edge than elsewhere in the nanotube. In the arm-chair edge,
however, it turns out that the most stable structure is that in which
the boron atom is placed at a site one bond away from the edge, 
i.e. site $\beta$ in Figure~\ref{fig:sites}(b), which turns out to be 0.58~eV
more stable than the structure obtained by placing the boron atom
directly at the edge (site $\alpha$). Nevertheless, boron will 
still preferentially locate
itself close to the edge rather than far away from it, as the energy
difference between the most stable configuration and that with boron
placed in site $\epsilon$ is 0.88~eV.
These findings can be easily rationalized: in the zig-zag case it is 
more favorable for boron to be at the edge than elsewhere in the tube
because in this way the number of carbon dangling bonds at the nanotube
edge is reduced, given that boron has one electron less than carbon.
In the arm-chair case, however, the same argument does not apply, because
the carbon atoms at the edge are able to pair up forming a stable
double bond. By placing a boron atom at the edge a double bond cannot form,
and the configuration that results is thus less stable than
configuration $\beta$. The same trends are evident from the calculations
involving the flat graphene edges, indicating that the curvature of 
the nanotubes does not play any significant part in determining 
the energetics of the boron-substituted structures.

\begin{table}
\caption{Energies of boron-substituted structures for zig-zag and
arm-chair nanotube and graphene edges. The energies are given in eV relative to
the most stable structure in each case.
\label{table:energies}}
\begin{center}
\begin{tabular}{ccccc}
bonds & (10,0) & zig-zag & (6,6) & arm-chair \\
from edge & nanotube & edge & nanotube & edge \\
\hline
0 & 0.00 & 0.00 & 0.58 & 0.48 \\
1 & 1.39 & 0.99 & 0.00 & 0.00 \\
2 & 0.57 & 0.67 & 0.60 & 0.66 \\
3 & 1.30 & 1.42 & 0.83 & 0.75 \\
6 & 1.41 & 1.35 & 0.88 & 0.88 \\
\end{tabular}
\end{center}
\end{table}

These static calculations illustrate how boron could play a role in
favoring zig-zag structures over arm-chair ones, as has been found
experimentally. Our findings are in very good agreement with the 
DFT plane-wave pseudo-potential results 
reported by Blase and coworkers~\cite{blase:terrones}. However, these results
do not give any explanation for the increased length of the nanotubes
synthesized in the presence of boron. Thus, in order to gain some insight
into how boron can assist the production of longer nanotubes, we have
performed a series of MD simulations of a (10,0) nanotube with varying
degrees of boron substitution at the edge. We have employed the same
structure as illustrated in Figure~\ref{fig:starting}, but placing 
four, six, eight and ten boron atoms at the tube edge, substituting as 
many carbon atoms in each case. Starting from such configurations, we have
performed MD simulations following the same pattern described above for
the undoped carbon nanotubes.
Let us comment briefly on the details of the dynamics in 
each case. In all instances, at the lowest temperature considered (1000~K),
the structures remain unaltered during the time spanned at this temperature
(10~ps); all hexagonal rings present initially maintain their identity 
during this time, and it is not until the system is run at a temperature of 
2000~K  that structural changes begin to appear.

\begin{figure}
\begin{minipage}[t]{6.5cm}
\begin{center}
\begin{center}
{\bf \large (a)}
\end{center}
\epsfxsize=6.5cm
\epsffile{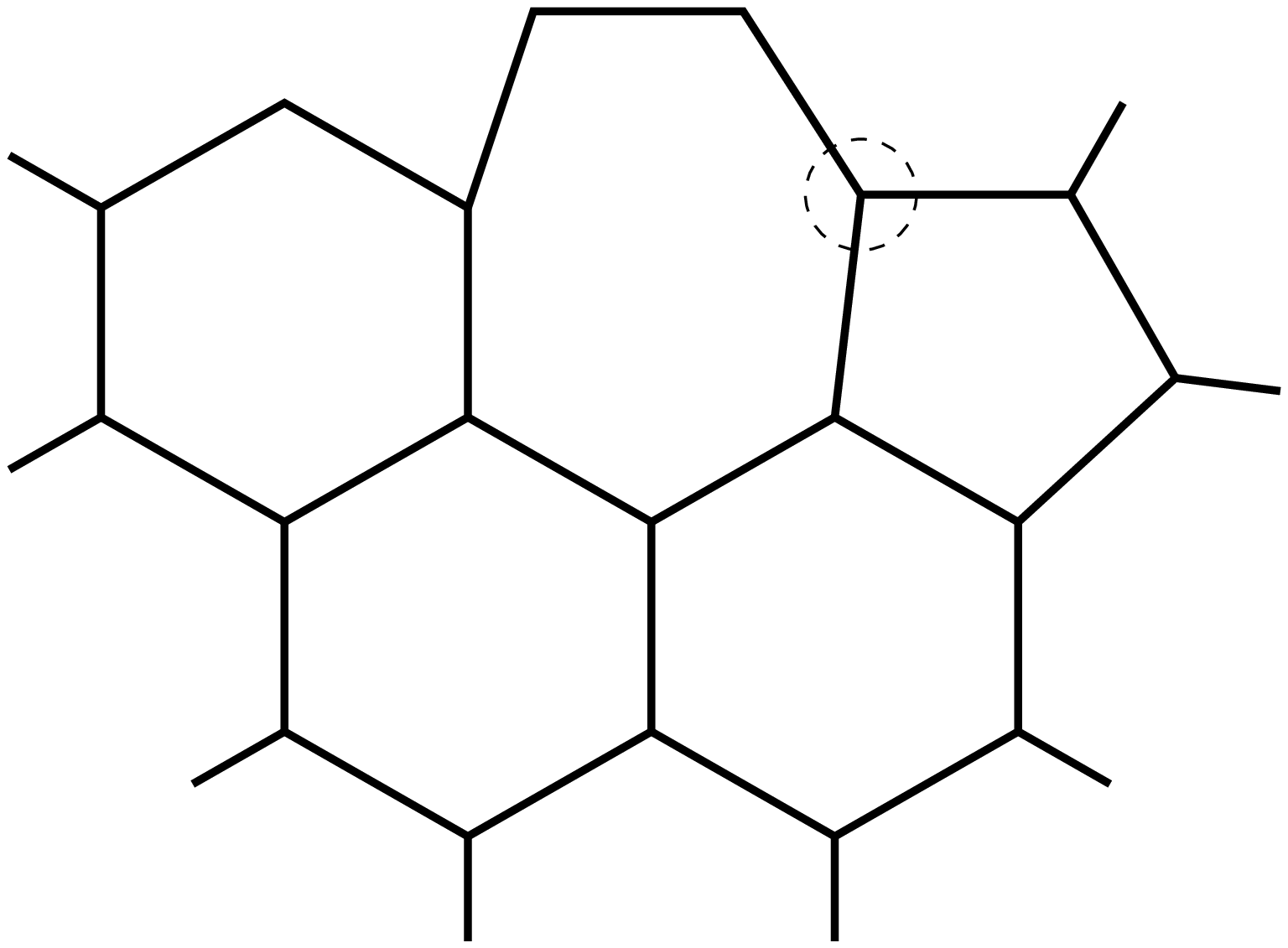}
\end{center}
\end{minipage}
\begin{minipage}[h]{6.5cm}
\begin{center}
{\bf \large (b)}
\end{center}
\epsfxsize=6.5cm
\epsffile{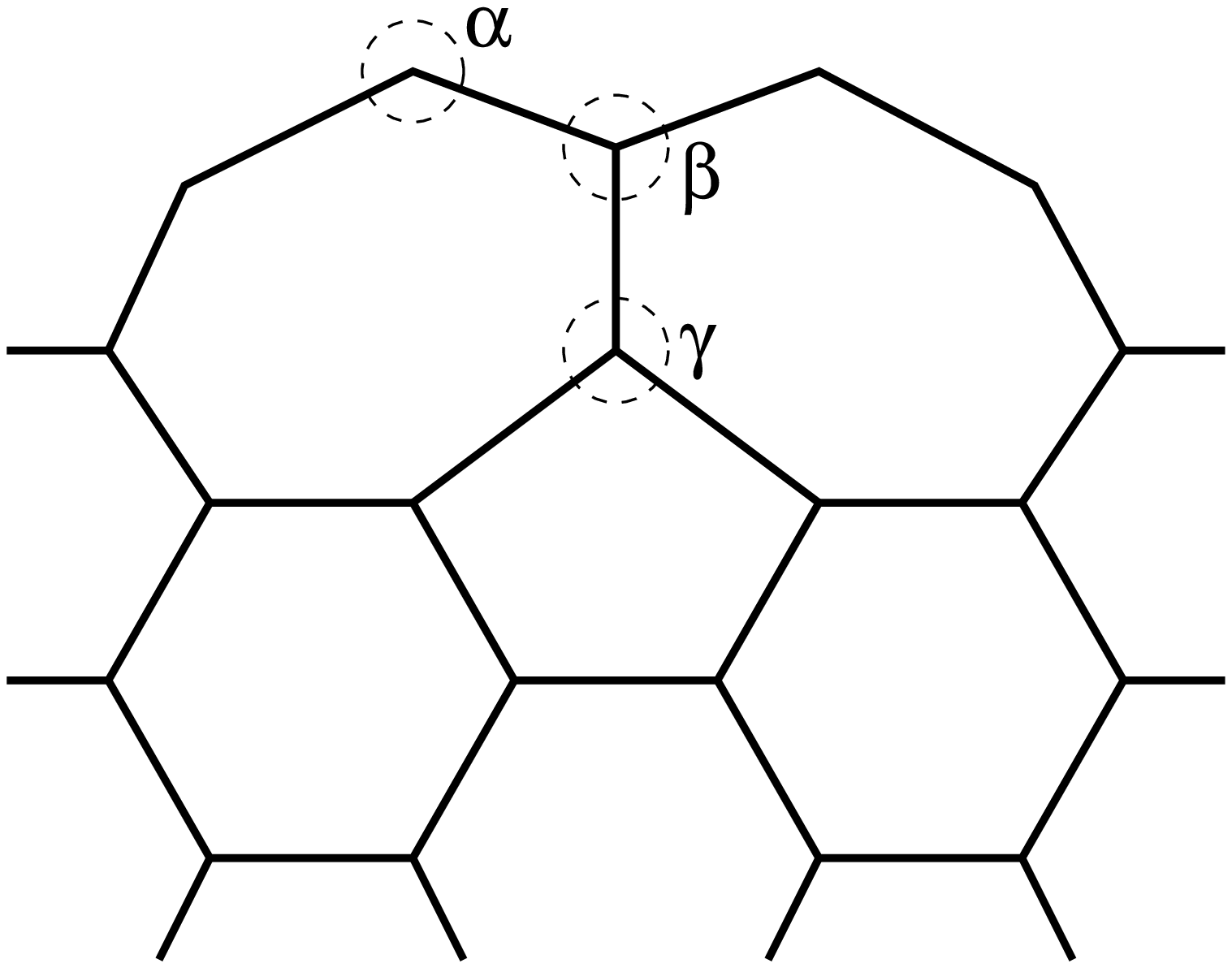}
\end{minipage}
\begin{minipage}[b]{6.5cm}
\begin{center}
{\bf \large (c)}
\leavevmode
\epsfxsize=6.5cm
\epsffile{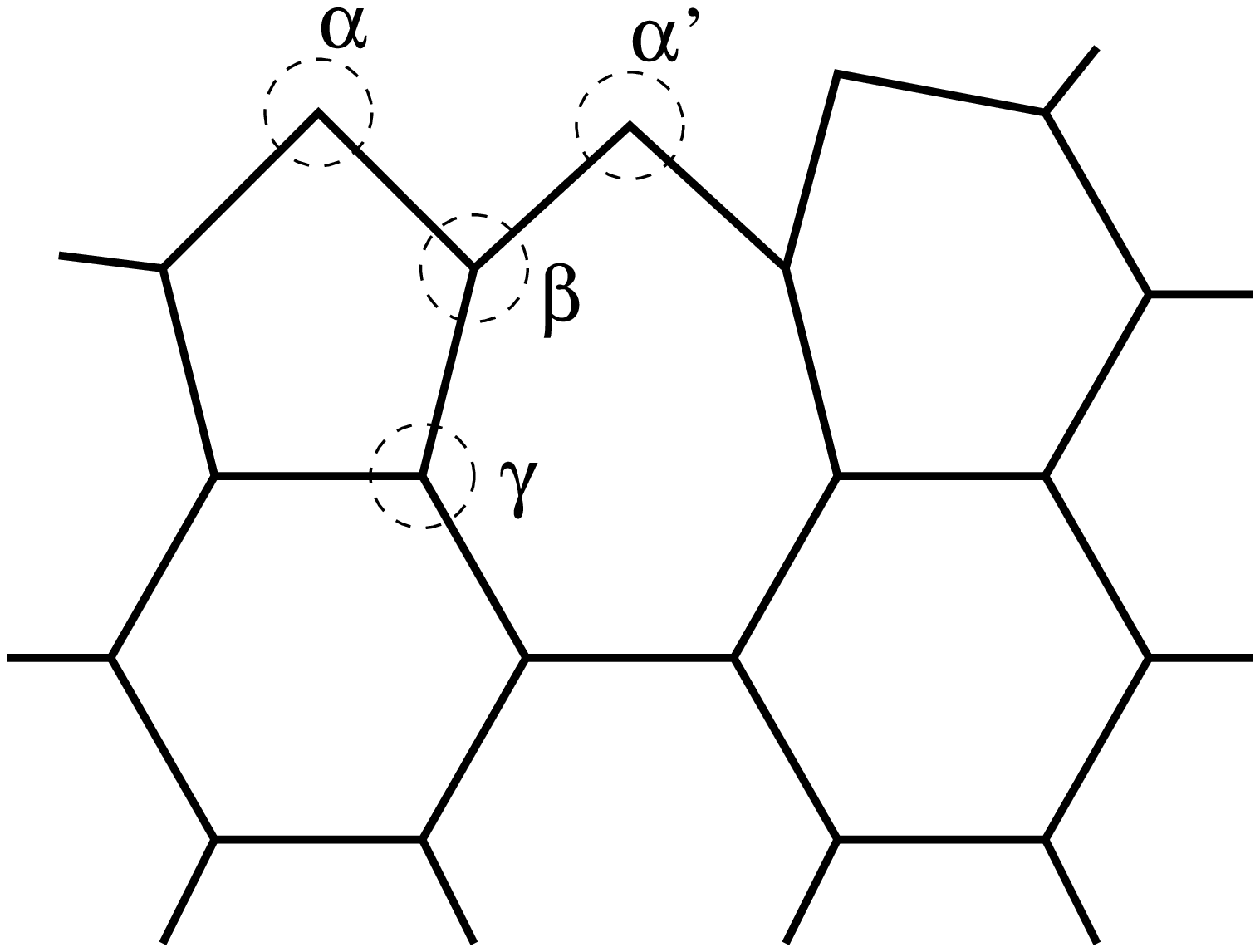}
\end{center}
\end{minipage}
\caption{Schematic view of (a) the (5/7)~edge complex observed during
the MD runs of the (10,0) nanotube (the circle indicates the position where
boron is found in the boron doped cases); (b) one of the the possible edge
reconstructions involving pentagons and heptagons in the arm-chair type
tubes, the (7/5/7) complex, and (c) the (5/7) arm-chair edge complex. The
circles indicate the substitutional sites considered in the relaxation
calculations (see text).}
\label{fig:5/7}
\end{figure}

	When only four boron atoms are present, after 10~ps of dynamics at
2000~K two (5/7)~pairs are formed at the tube edge. The 
structure of these (5/7)~pairs is such that the common edge
between both rings always incorporates a boron atom [see 
Figure~\ref{fig:5/7}(a)]. This is a structural feature that repeats itself
frequently in the other cases considered, as will be seen below. At this
temperature, a transient $\mbox{C}_4$ ring, and a $\mbox{B}_2\mbox{C}_6$
octagonal ring are also seen. When the temperature is raised further
to 2500~K, the boron containing pentagons still survive, but the heptagonal
rings break, and carbon chains linked to the pentagons are formed. These
chains, of length equal to two or three bonds, oscillate widely, eventually
linking with similar chains or unsaturated bonds along the open edge,
thus initiating the tube closure. The tube is closed after approximately
8.5~ps at this temperature.  Given the low concentration of boron at the tip,
the dynamics observed here is rather similar to that already described for
the pure carbon case.

	When the number of boron atoms is increased to 6, at the temperature
of 2000~K two (5/7)~complexes like those observed in the 
four-boron atom case are formed, but the structure is otherwise unaltered.
These boron-containing (5/7)~pairs are remarkably stable;
they remain intact even after 10~ps at 2500~K. In this particular case,
it is necessary to rise the temperature further, to 3000~K, for the tube
closure to occur. This happens only after 8~ps of dynamics at this temperature,
which is indicative of a higher stability of the open edge when compared
to the tube with only four boron atoms.
In the case of the nanotube doped with eight boron atoms, the 
structure develops two (5/7)~pairs within 10~ps at 2000~K, much like in the
previous cases. However, once the temperature is elevated to 2500~K, the
structure closes completely within 7.5~ps. This result appears to contradict
the thesis that increasing the amount of boron at the tube edge delays the
closure process, but in fact this is not necessarily so, as will be argued
below.

	Finally, let us consider the case of the tube with ten boron atoms.
Like in all the cases analyzed earlier, structural changes at the open edge
begin to appear at a temperature of 2000~K, at which a boron-containing
(5/7)~pair is formed. Three such complexes can be seen after 10~ps at
2500~K, but the structure is otherwise unchanged. It is necessary to reach
a temperature of 3000~K and monitor the dynamics of the system for 11~ps
at this temperature before the tube can be said to be completely closed.
We note that Blase and coworkers~\cite{blase:terrones}, who have also performed 
MD simulations of this particular system using DFT with a basis set
of plane-waves and the pseudo-potential approximation up to temperatures
of 2500~K for 10~ps, failed to observe the closure. 
The results reported here indicate
that, at such a temperature, it is most likely that the closure would not
occur below 10~ps, as it required 11~ps at 3000~K to observe the closure
in the simulations that we have performed.

\begin{figure}
\begin{minipage}[t]{3.5cm}
\begin{center}
{\bf \large (a)}
\end{center}
\epsfxsize=3.5cm
\epsffile{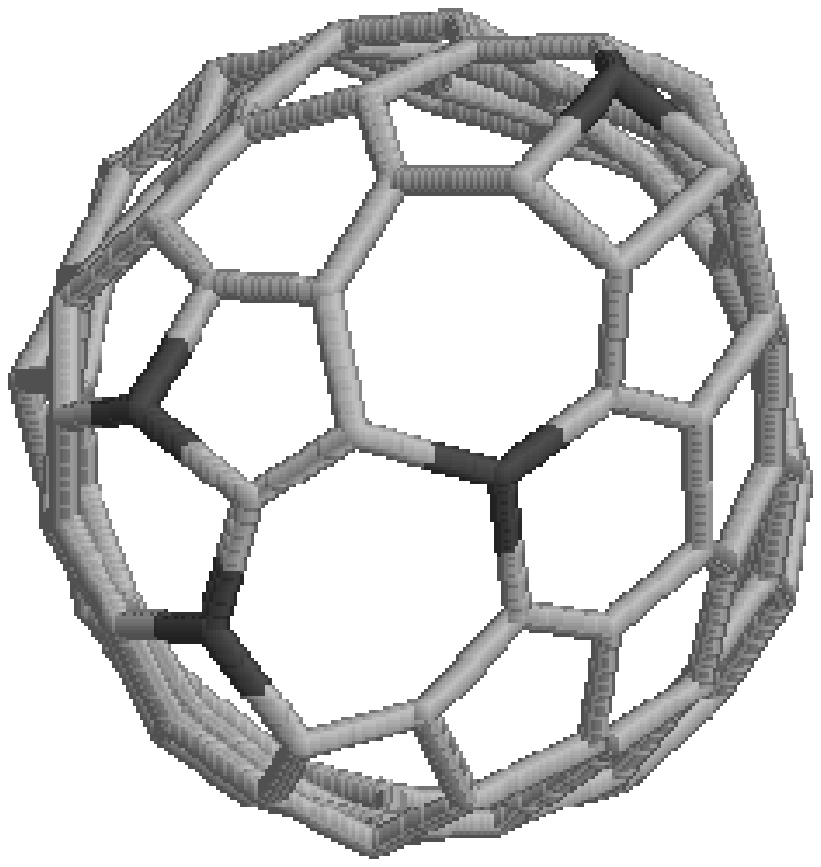}
\end{minipage}
\begin{minipage}[t]{3.5cm}
\begin{center}
{\bf \large (b)}
\end{center}
\epsfxsize=3.5cm
\epsffile{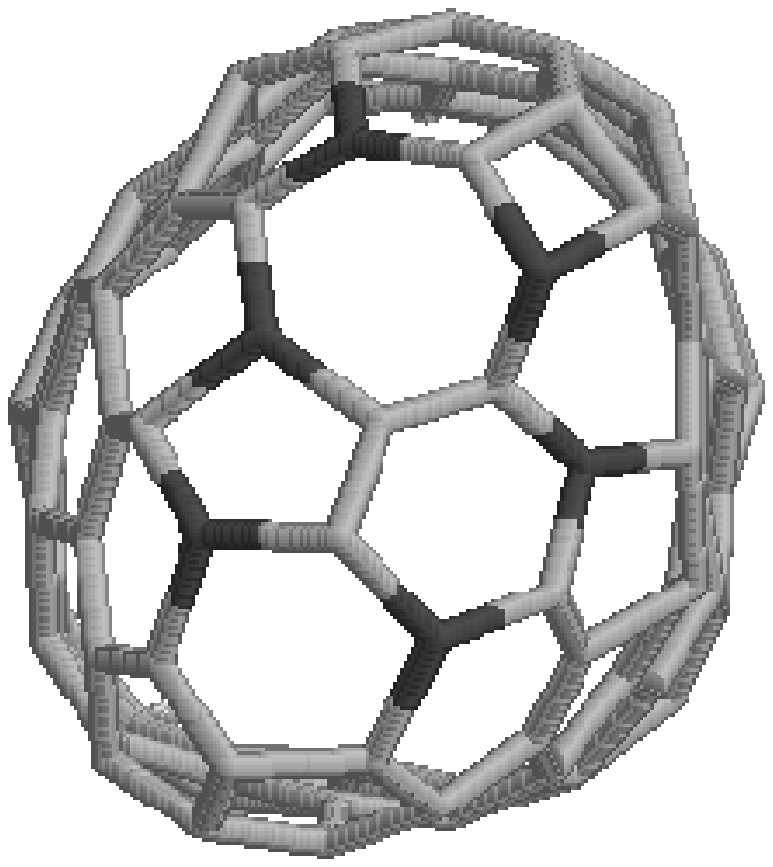}
\end{minipage}
\begin{minipage}[b]{3.5cm}
\begin{center}
{\bf \large (c)}
\end{center}
\epsfxsize=3.5cm
\epsffile{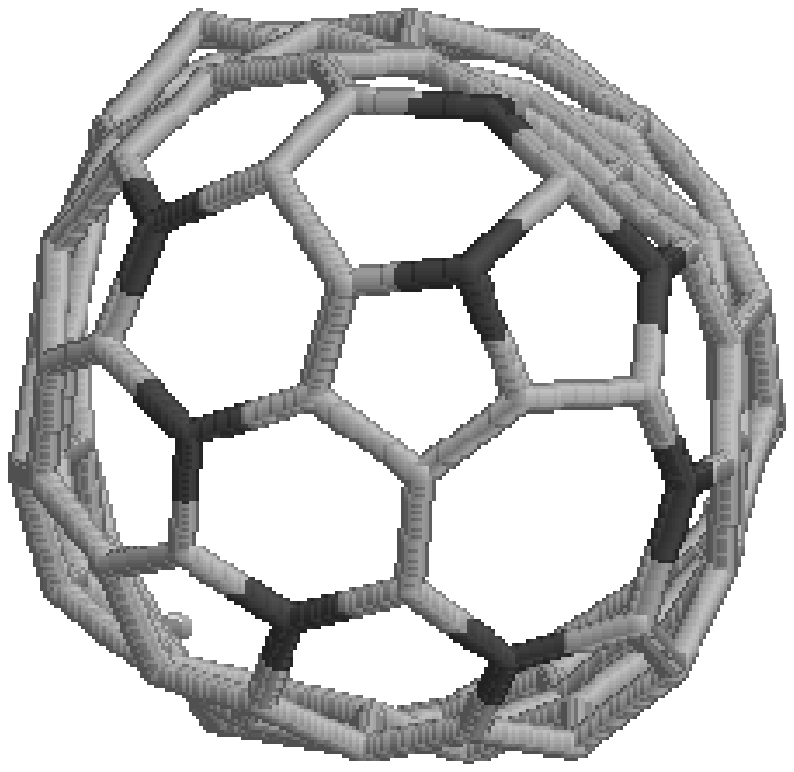}
\end{minipage}
\begin{minipage}[b]{3.5cm}
\begin{center}
{\bf \large (d)}
\end{center}
\epsfxsize=3.5cm
\epsffile{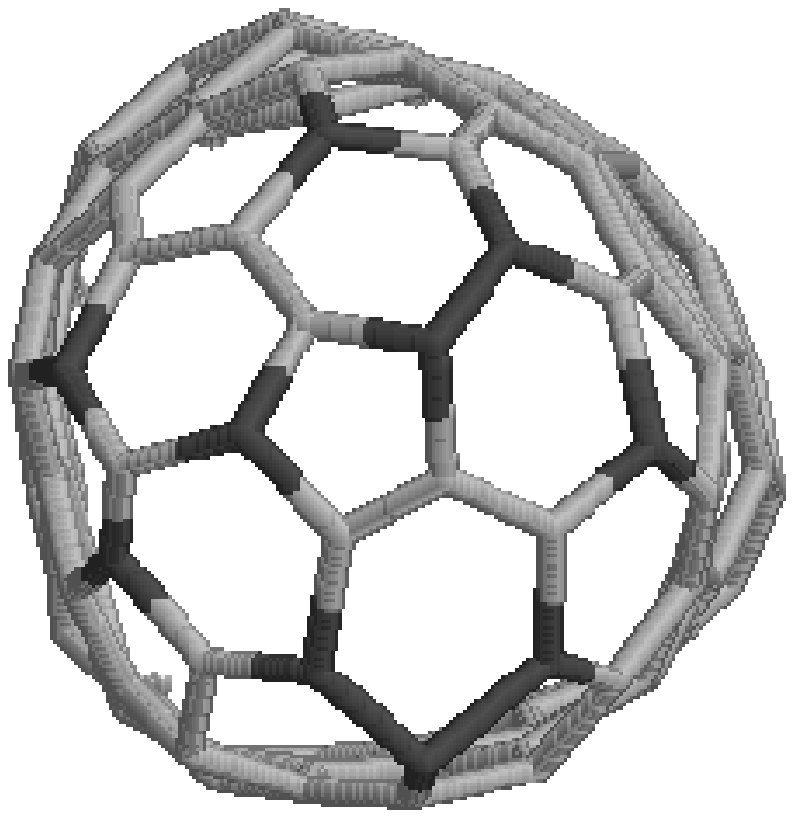}
\end{minipage}
\caption{The dome structures of the closed boron doped nanotubes arrived
at during the MD simulations: (a)~nanotube with four boron atoms,
(b)~nanotube with six boron atoms, (c)~nanotube with eight boron atoms
and (d)~with ten boron atoms.}
\label{fig:boron_dynamics}
\end{figure}

	Figure~\ref{fig:boron_dynamics} illustrates the structures obtained at
the end of the MD simulations described above. As can be seen, regardless
of the amount of boron present initially at the tube edge, the final 
structures are invariably closed, but as seen in the discussion, there are
notable differences observed during the dynamics in each case. That the 
nanotubes evolve in such a way as to achieve a closed structure is, after all,
not so surprising, given that the closed structures, with all atoms connected
in an $\mbox{\em sp\/}^2$ network are always more stable than the 
corresponding open structures. For example, the structure illustrated
in Figure~\ref{fig:boron_dynamics}(d), once relaxed at 0~K, is 13.7~eV
more stable than the corresponding open structure (a figure to be 
compared with an energy difference of 17~eV, obtained for the pure carbon
case using the same two structures). Nevertheless, as can be seen from the
discussion above, the details of the closure 
process vary according to the amount of boron present at
the edge. 

	Our MD results seem to indicate that a delay of the tube closure
occurs when boron is present in the system. However, it is difficult to 
establish this beyond doubt purely on the basis of a small number of 
MD trajectories, because at each temperature and boron composition there will
be a distribution of possible closure times (depending on the initial 
conditions) for a given type of nanotube. To get a clear picture of how the
closure time distribution changes as the amount of boron is increased
would require a very large sample of MD simulations, much larger than 
would be practical to carry out given the computational costs involved.
Our simulations are clearly insufficient in number to reveal 
unambiguously the changing trends in closure times with increasing boron
content, which could explain the apparently contradicting result 
obtained for the tube with eight boron atoms, which closes faster and
at a lower temperature than the tube with only six boron atoms. 
Nevertheless, these simulations are sufficient to clarify the role
played by boron when present at the open edge of a nanotube, and are
thus extremely revealing. They indicate that 
before the tube closure can begin to occur, (5/7)~complexes form at the edge.
This is true of both the pure carbon and the boron doped tubes. Under
the effect of the rapid thermal motion the heptagons seem to be more 
prone to breaking, and the chains of atoms thus formed initiate or 
can participate in the tube closure. Interestingly, when boron is present,
it seems to be always associated with these structures once they are
formed, in the manner depicted in Figure~\ref{fig:5/7}(a). 

	These observations have lead us to perform a new set of static
structural relaxation calculations, in which we investigate the stability
of these edge (5/7)~complexes in both (n,0) and (n,n) nanotubes. We have
compared the energy of a normal zig-zag edge structure, such as that
shown in Figure~\ref{fig:starting} with that of a zig-zag edge
containing a (5/7)~pair, both in the pure carbon case and in the case in
which the (5/7)~pair contains a boron atom situated between the pentagon
and heptagon at the edge [Figure~\ref{fig:5/7}(a)]. It turns out that 
a (5/7)~pair placed
at the edge of a pure (10,0) tube lowers the total energy by about 
0.56~eV. When boron is present between the pentagon and heptagon, the 
energy gain is slightly larger, 0.67~eV, according to our calculations. This 
seems to indicate that the formation of such boron containing 
edge complexes stabilizes the open edge, and consequently delays its 
closure. 

	A (5/7)~complex is also possible at an arm-chair nanotube edge,
where it is also possible to construct a more complicated edge structure,
involving a pentagon and two heptagons, which we call a (7/5/7)~edge
complex; both structures are illustrated schematically in
Figure~\ref{fig:5/7}(b) and \ref{fig:5/7}(c).  Structure~\ref{fig:5/7}(b) is 
reminiscent of the ring-defect structure discussed by
Crespi {\em et al.\/}~\cite{crespi}. We have investigated the 
stability of these, both in the pure carbon case and when boron is 
present in the different possible substitutional sites.  The
results from these calculations are listed in 
Table~\ref{table:arm_chair_edge_results}. It turns out that the only 
reconstructed arm-chair edge which is more stable than the structure~$\beta$
shown in Figure~\ref{fig:sites}(b) is that illustrated in
Figure~\ref{fig:5/7}(b), with the boron atom placed between
the two heptagonal rings, but not forming part of the pentagon ({\em i.e.\/}
site~$\beta$). Even so, the energy difference is only small (0.33~eV), 
certainly smaller than the energy gain obtained by the reconstruction 
in the zig-zag edge. 

\begin{table}
\caption{Energies of boron-substituted reconstructed edges for the
(6,6) nanotube. The labeling of the edges corresponds to that of
Figure~\ref{fig:5/7} (b) and (c). The energies are given in eV relative to
the most stable unreconstructed substituted arm-chair edge structure.}
\label{table:arm_chair_edge_results}
\begin{center}
\begin{tabular}{ccc}
Position & Structure (b) & Structure (c) \\
\hline
$\alpha$ & 1.10 & 3.40 \\
$\beta$ & -0.33 & 3.17 \\
$\gamma$ & 0.46 & 2.67 \\
$\alpha\prime$ & -- & 2.64 \\
\end{tabular}
\end{center}
\end{table}

	The results obtained from these structural relaxation calculations
provide a possible mechanism for explaining both the observed preference for
the zig-zag structure and the improved aspect ratio  of nanotubes
synthesized in the presence of boron. The reconstruction of a zig-zag
edge in which boron is present from the saw-tooth structure illustrated
in Figure~\ref{fig:starting} to one containing a (5/7)~pair is 
exothermic, resulting in a significant stabilization of the nanotube.
In the case of an undoped nanotube edge, this reconstruction also lowers
the energy, but by a smaller amount. Note also that, while similar
reconstructions are possible in the case of a boron doped arm-chair nanotube,
only one such reconstructions lowers the energy with respect to the 
unreconstructed edge, and only by a small amount. Boron, thus,
is seen to significantly stabilize zig-zag edges compared to arm-chair
ones and we expect the former tubes to grow better; furthermore the edge 
(5/7) relaxation mechanism we have found is 
only effective in zig-zag edges, a fact that contributes to favor this
type of tube even further.

\section{Conclusions}
\label{sec:concs}

	We have performed an extensive theoretical study of the effects of 
the presence of varying amounts of boron in the edge of growing nanotubes
using TB MD and structural relaxation calculations. In spite of the 
approximate nature of the TB model used here, it gives results that
compare quite well with the available FP data from the work of 
Blase {\em et al.\/}~\cite{blase:terrones}.

	We have shown that both carbon and boron-doped nanotubes spontaneously
close in a time scale of a few ($\leq 20$)~ps in a temperature range of
2500-3000~K. In the case of boron-doped tubes the MD simulations reveal a 
pattern of longer closure times as the amount of boron at the tip of the
nanotube is increased. The presence of boron in the proximity of the 
tube edge has a stabilizing effect, but this effect is larger in the case
of zig-zag nanotubes than in the case of arm-chair ones. Furthermore, we
have shown that in (n,0) edges a reconstruction of the edge involving
(5/7)~pairs stabilizes further the open edge. Although similar reconstructions
are topologically possible in the case of (n,n) edges they do not result in
any significant stabilization of this type of edge.

	These results contribute to a better understanding of the 
experimental observations~\cite{redlich,mauro_apl,blase:terrones} that boron
accumulates at the nanotube tips, that it favors zig-zag nanotubes over
other structures, and that the nanotubes synthesized in the presence of 
boron have a larger aspect ratio than other nanotubes.

* to whom correspondence should be addressed; emai: ehe@icmab.es

\acknowledgements

E.H. wishes to thank X.~Blase, M.~Terrones, N.~Grobert, 
W.K.~Hsu, H.~Terrones and M.J.~L\'{o}pez
for enlightening discussions. E.H. and P.O. thank the EU for financial
support under project SATURN (IST-1999-10593).
A.R and J.A.A ackowledge support by the DGES (Grant: PB98-0345),
JCyL (Grant: VA28/99), and EU TMR
NAMITECH project (ERBFMRX-CT96-0067 (DG12-MITH)). 
I.B. acknowledges support 
by the DGES (SAB 1995-0670P) of Spain during the sabbatical stay 
at the University of Valladolid, by the DFG (Deutsche 
Forschungsgemeinschaft Project SPP-Polyeder), and finally by
the Fonds der Chemischen Industrie.
We acknowledge the C$^4$ (Centre de Computaci\'o i Comunicacions de Catalunya)
and CEPBA (European Centre for Parallelism of Barcelona) 
for the use of their computer facilities.

\end{document}